\documentclass[a4paper,11pt]{article}

\usepackage[T1]{fontenc} 
\usepackage{float} 

\makeatletter
\gdef\@fpheader{ }
\gdef\@journal{ }
\makeatother
\RequirePackage{amsmath}
\RequirePackage{amssymb}
\RequirePackage{epsfig}
\RequirePackage{graphicx}
\RequirePackage[numbers,sort&compress]{natbib}
\RequirePackage{color}
\RequirePackage[colorlinks=true
,urlcolor=blue
,anchorcolor=blue
,citecolor=blue
,filecolor=blue
,linkcolor=blue
,menucolor=blue
,pagecolor=blue
,linktocpage=true
,pdfproducer=medialab
,pdfa=true
]{hyperref}

\newif\ifnotoc\notocfalse
\newif\ifemailadd\emailaddfalse
\newif\iftoccontinuous\toccontinuousfalse
\makeatletter
\def\@subheader{\@empty}
\def\@keywords{\@empty}
\def\@abstract{\@empty}
\def\@xtum{\@empty}
\def\@dedicated{\@empty}
\def\@arxivnumber{\@empty}
\def\@collaboration{\@empty}
\def\@collaborationImg{\@empty}
\def\@proceeding{\@empty}
\def\@preprint{\@empty}

\newcommand{\subheader}[1]{\gdef\@subheader{#1}}
\newcommand{\keywords}[1]{\if!\@keywords!\gdef\@keywords{#1}\else%
\PackageWarningNoLine{\jname}{Keywords already defined.\MessageBreak Ignoring last definition.}\fi}
\renewcommand{\abstract}[1]{\gdef\@abstract{#1}}
\newcommand{\dedicated}[1]{\gdef\@dedicated{#1}}
\newcommand{\arxivnumber}[1]{\gdef\@arxivnumber{#1}}
\newcommand{\proceeding}[1]{\gdef\@proceeding{#1}}
\newcommand{\xtumfont}[1]{\textsc{#1}}
\newcommand{\correctionref}[3]{\gdef\@xtum{\xtumfont{#1} \href{#2}{#3}}}
\newcommand\jname{JHEP}

\newcommand\preprint[1]{\gdef\@preprint{\hfill #1}}

\makeatother

\newcommand\note[2][]{%
\if!#1!%
\stepcounter{footnote}\footnotetext{#2}%
\else%
{\renewcommand\thefootnote{#1}%
\footnotetext{#2}}%
\fi}

\makeatletter
\newtoks\auth@toks
\renewcommand{\author}[2][]{%
  \if!#1!%
    \auth@toks=\expandafter{\the\auth@toks#2\ }%
  \else
    \auth@toks=\expandafter{\the\auth@toks#2$^{#1}$\ }%
  \fi
}
\makeatother
\makeatletter
\newtoks\affil@toks\newif\ifaffil\affilfalse
\newcommand{\affiliation}[2][]{%
\affiltrue
  \if!#1!%
    \affil@toks=\expandafter{\the\affil@toks{\item[]#2}}%
  \else
    \affil@toks=\expandafter{\the\affil@toks{\item[$^{#1}$]#2}}%
  \fi
}
\makeatother
\makeatletter
\newtoks\email@toks\newcounter{email@counter}%
\newcommand{\emailAdd}[1]{%
\emailaddtrue%
\ifnum\theemail@counter>0\email@toks=\expandafter{\the\email@toks, \@email{#1}}%
\else\email@toks=\expandafter{\the\email@toks\@email{#1}}%
\fi\stepcounter{email@counter}}
\newcommand{\@email}[1]{\href{mailto:#1}{\tt #1}}
\makeatother

\makeatletter
\newcommand*\collaboration[1]{\gdef\@collaboration{#1}}
\newcommand*\collaborationImg[2][]{\gdef\@collaborationImg{#2}}
\makeatletter
\newcommand\afterLogoSpace{\smallskip}
\newcommand\afterSubheaderSpace{\vskip3pt plus 2pt minus 1pt}
\newcommand\afterProceedingsSpace{\vskip21pt plus0.4fil minus15pt}
\newcommand\afterTitleSpace{\vskip23pt plus0.06fil minus13pt}
\newcommand\afterRuleSpace{\vskip23pt plus0.06fil minus13pt}
\newcommand\afterCollaborationSpace{\vskip3pt plus 2pt minus 1pt}
\newcommand\afterCollaborationImgSpace{\vskip3pt plus 2pt minus 1pt}
\newcommand\afterAuthorSpace{\vskip5pt plus4pt minus4pt}
\newcommand\afterAffiliationSpace{\vskip3pt plus3pt}
\newcommand\afterEmailSpace{\vskip16pt plus9pt minus10pt\filbreak}
\newcommand\afterXtumSpace{\par\bigskip}
\newcommand\afterAbstractSpace{\vskip16pt plus9pt minus13pt}
\newcommand\afterKeywordsSpace{\vskip16pt plus9pt minus13pt}
\newcommand\afterArxivSpace{\vskip3pt plus0.01fil minus10pt}
\newcommand\afterDedicatedSpace{\vskip0pt plus0.01fil}
\newcommand\afterTocSpace{\bigskip\medskip}
\newcommand\afterTocRuleSpace{\bigskip\bigskip}
\newlength{\affiliationsSep}
\newcommand\beforetochook{\pagestyle{myplain}\pagenumbering{roman}}

\DeclareFixedFont\trfont{OT1}{phv}{b}{sc}{11}

\renewcommand\maketitle{
\pagestyle{empty}
\thispagestyle{titlepage}
\setcounter{page}{0}
\noindent{\small\scshape\@fpheader}\@preprint\par

\afterLogoSpace
\if!\@subheader!\else\noindent{\trfont{\@subheader}}\fi
\afterSubheaderSpace
\if!\@proceeding!\else\noindent{\sc\@proceeding}\fi
\afterProceedingsSpace
{\LARGE\flushleft\sffamily\bfseries\@title\par}
\afterTitleSpace
\hrule height 1.5\p@%
\afterRuleSpace
\if!\@collaboration!\else
{\Large\bfseries\sffamily\raggedright\@collaboration}\par
\afterCollaborationSpace
\fi
\if!\@collaborationImg!\else
{\normalsize\bfseries\sffamily\raggedright\@collaborationImg}\par
\afterCollaborationImgSpace
\fi
{\bfseries\raggedright\sffamily\the\auth@toks\par}
\afterAuthorSpace
\ifaffil\begin{list}{}{%
\setlength{\leftmargin}{0.28cm}%
\setlength{\labelsep}{0pt}%
\setlength{\itemsep}{\affiliationsSep}%
\setlength{\topsep}{-\parskip}}
\itshape\small%
\the\affil@toks
\end{list}\fi
\afterAffiliationSpace
\ifemailadd 
\noindent\hspace{0.28cm}\begin{minipage}[l]{.9\textwidth}
\begin{flushleft}
\textit{E-mail:} \the\email@toks
\end{flushleft}
\end{minipage}
\else 
\PackageWarningNoLine{\jname}{E-mails are missing.\MessageBreak Plese use \protect\emailAdd\space macro to provide e-mails.}
\fi
\afterEmailSpace
\if!\@xtum!\else\noindent{\@xtum}\afterXtumSpace\fi
\if!\@abstract!\else\noindent{\renewcommand\baselinestretch{.9}\textsc{Abstract:}}\ \@abstract\afterAbstractSpace\fi
\if!\@keywords!\else\noindent{\textsc{Keywords:}} \@keywords\afterKeywordsSpace\fi
\if!\@arxivnumber!\else\noindent{\textsc{ArXiv ePrint:}} \href{http://arxiv.org/abs/\@arxivnumber}{\@arxivnumber}\afterArxivSpace\fi
\if!\@dedicated!\else\vbox{\small\it\raggedleft\@dedicated}\afterDedicatedSpace\fi
\ifnotoc\else
\iftoccontinuous\else\newpage\fi
\beforetochook\hrule
\tableofcontents
\afterTocSpace
\hrule
\afterTocRuleSpace
\fi
\setcounter{footnote}{0}
\pagestyle{myplain}\pagenumbering{arabic}
} 

\renewcommand{\baselinestretch}{1.1}\normalsize
\divide\@tempdima\baselineskip
\@tempcnta=\@tempdima
\skip\@mpfootins = \skip\footins
\renewcommand{\@dotsep}{10000}

\newcommand\ps@myplain{
\pagenumbering{arabic}
\renewcommand\@oddfoot{\hfill-- \thepage\ --\hfill}
\renewcommand\@oddhead{}}
\let\ps@plain=\ps@myplain

\newcommand\ps@titlepage{\renewcommand\@oddfoot{}\renewcommand\@oddhead{}}

\numberwithin{equation}{section}

\renewcommand\section{\@startsection{section}{1}{\z@}%
                                   {-3.5ex \@plus -1.3ex \@minus -.7ex}%
                                   {2.3ex \@plus.4ex \@minus .4ex}%
                                   {\normalfont\large\bfseries}}
\renewcommand\subsection{\@startsection{subsection}{2}{\z@}%
                                   {-2.3ex\@plus -1ex \@minus -.5ex}%
                                   {1.2ex \@plus .3ex \@minus .3ex}%
                                   {\normalfont\normalsize\bfseries}}
\renewcommand\subsubsection{\@startsection{subsubsection}{3}{\z@}%
                                   {-2.3ex\@plus -1ex \@minus -.5ex}%
                                   {1ex \@plus .2ex \@minus .2ex}%
                                   {\normalfont\normalsize\bfseries}}
\renewcommand\paragraph{\@startsection{paragraph}{4}{\z@}%
                                   {1.75ex \@plus1ex \@minus.2ex}%
                                   {-1em}%
                                   {\normalfont\normalsize\bfseries}}
\renewcommand\subparagraph{\@startsection{subparagraph}{5}{\parindent}%
                                   {1.75ex \@plus1ex \@minus .2ex}%
                                   {-1em}%
                                   {\normalfont\normalsize\bfseries}}

\def\fnum@figure{\textbf{\figurename\nobreakspace\thefigure}}
\def\fnum@table{\textbf{\tablename\nobreakspace\thetable}}

\long\def\@makecaption#1#2{%
  \vskip\abovecaptionskip
  \sbox\@tempboxa{\small #1. #2}%
  \ifdim \wd\@tempboxa >\hsize
    \small #1. #2\par
  \else
    \global \@minipagefalse
    \hb@xt@\hsize{\hfil\box\@tempboxa\hfil}%
  \fi
  \vskip\belowcaptionskip}

\renewenvironment{thebibliography}[1]{%
\begin{oldthebibliography}{#1}%
\small%
\raggedright%
\setlength{\itemsep}{5pt plus 0.2ex minus 0.05ex}%
}%
{%
\end{oldthebibliography}%
}

\begin{document}


\title{\boldmath To Simulate the Spread of Infectious Diseases by the Random Matrix}


\author[a]{Ting Wang,}
\author[a]{Gui-Yun Li,}\note{Guiyun Li and Ting Wang contributed equivalently to this work.}
\author[a]{Xin-Hui Li,}
\author[a]{Chi-Chun Zhou,}\note{zhouchichun@dali.edu.cn. Corresponding author.}
\author[a]{Yuan-Yuan Wang,}
\author[a]{Li-Juan Li,}
\author[a]{and Yan-Ting Yang}

\affiliation[a]{School of Engineering, Dali University, Dali, Yunnan 671003, PR China}










\abstract{The main aim to build models capable 
of simulating the spreading of infectious diseases is to control them.
And along this way, the key to find the optimal strategy 
for disease control is to obtain a 
large number of simulations of disease transitions under different scenarios.
Therefore, the models that can simulate the spreading  of diseases under scenarios closer 
to the reality and are with high efficiency are preferred.
In the realistic social networks, 
the random contact, including contacts between people in the public places and
the public transits, becomes the important access for the spreading of infectious 
diseases.
In this paper, a model can efficiently simulate the spreading of infectious 
diseases under random contacts is proposed.
In this approach, the random contact between people is characterized by 
the random matrix with elements randomly generated and the spread of the diseases 
is simulated by the Markov process.
We report an interesting property of the proposed model:
the main indicators of the spreading of the diseases such as the death rate are invariant 
of the size of the population. 
Therefore, representative simulations can be conducted on models consist of small number of 
populations.
The main advantage of this model is that it can easily simulate
the spreading of diseases under more realistic scenarios 
and thus is able to give a large number of simulations needed for the searching of the 
optimal control strategy. 
Based on this work, the reinforcement learning will be introduced 
to give the optimal control strategy in the following work.}


\maketitle
\flushbottom

\section{Introduction}

Infectious diseases, especially malignant infectious diseases 
that can spread on a large scale and 
cause large-scale deaths, seriously threaten the safety of 
individual lives and social orders \cite{2017Massively}, 
and are major problems in modern public health events.
For example, the Black Death in the mid-14th century swept across Europe, 
killing about $1/3$ of the European population at that time \cite{dewitte2008selectivity}.
The SARS, outbreaks at 2003, is estimated to 
have cost the global economy more than $50$ billion dollars worth of 
damages \cite{fernandes2020economic}.
The 2019 novel coronavirus pneumonia (COVID-19)
has killed more than $6$ million people, the epidemic is still happening locally, and
the death toll will continue to rise. 
Moreover, they cause a lot of economic losses 
making the world economy fall into a downturn.

Establishing a model capable of simulating the spreading of infectious 
diseases is important to predicting the trend of infectious diseases 
and providing theoretical guidance for adopting the disease prevention measures.
Along this way, the key to find the optimal strategy 
for disease control is to obtain a 
large number of simulations of disease transitions under different scenarios closer 
to the reality.
Therefore, we need models that can efficiently simulate the spreading of 
diseases under realistic scenarios.

Researches on infectious disease models can be traced back to 1760, 
when Dutch physicist Bernoull uses mathematical models 
to study the effects of vaccinia vaccines on 
the spreading of smallpox \cite{2002Daniel}.
In 1906, Hamer used a discrete model to study 
the recurrent epidemic of measles \cite{1906The}.
In 1911, Ross used differential equations to study 
the transmission of malaria between mosquitoes and people, 
proving that when the number of mosquitoes is reduced 
below a critical value, the outbreak of malaria can be controlled \cite{ross1911prevention}.

Recently, the infectious disease models can be divided into three types.
(1) The methods based on differential equations.
In this approach, the dynamic process of disease transmission 
is simulated by setting different compartments to represent 
people in different disease states \cite{diekmann1995legacy,breda2012formulation}.
For example, the SIR model with $S$, $I$, and $R$ the susceptible, infectious and removed
populations is a typical compartment model. The disease transition is simulated by 
introducing the ordinary equations, $dS/dt=-\beta SI$, $dI/dt=\beta SI-\gamma I$, 
and $dR/dt=\gamma I$, where parameters $\beta$ and $\gamma$ stand for the infection
rate and the recovery rate respectively \cite{maki2013infectious}.
In order to simulate the spreading of diseases under more realistic scenarios, 
various compartment models are proposed.
For examples, the SEIR model \cite{eyaran2019modelling} (with $S$, $E$, $I$, and $R$ the susceptible, exposed, infected, 
and recover respectively), the 
SIRS model \cite{eyaran2019modelling} (with $S$, $I$, $R$, and $S$ the susceptible,
infectious, recovered, and susceptible respectively), and the model considers asymptomatic carriers 
\cite{chisholm2018implications} are 
proposed, where more classes or compartments according to the epidemiological 
status are considered. 
Beyond the deterministic compartment models, the stochastic compartment models
are introduced to simulate the stochastic factors 
\cite{gray2011stochastic,allen2017primer,maki2013infectious,islam2020mathematical}
and other types of differential equations, such as the 
delay differential equation \cite{huang2010lyapunov,eyaran2019modelling},
nonlinear Volterra integral equations \cite{greenhalgh2021generalized},
and the fractional differential equation \cite{islam2020mathematical},
are introduced into the compartment models.
The differential equation based compartment models are indeed able to 
simulate various kinds of infectious diseases 
\cite{meszaros2020direct,mukandavire2015modeling,
acevedo2015spatial,zheng2014modeling}
and evaluate the effectiveness of the control measures \cite{buchwald2020infectious,nadim2020occurrence}
however, they ignore the individual differences \cite{islam2020mathematical}, are unable to account for 
disease's true infectious period distribution \cite{greenhalgh2021generalized}, and are inflexible 
to simulate realistic scenarios such as 
random contact in the public places.

(2) The methods based on the cellular automata.
Unlike the compartment models, the cellular automata is individual–based–model, 
where each individual of the population is represented by a cell of 
the automata, and the transmission dynamic is simulated
by setting a set of simple rules that encode the
individual behaviour \cite{lopez2014addressing}.
The model based on cellular automata can simulate more realistic scenarios.
For example, it can simulate the impact of lockdown, 
migration and vaccination on COVID-19 dynamics 
\cite{jithesh2021model}, consider social community with varying
sex ratio, age structure, population movement, incubation and treatment period,
immunity \cite{dai2020modeling}, and address the population heterogeneity and
distribution in epidemics models \cite{lopez2014addressing,holko2016epidemiological}.
Beyond the conventional cellular automata models,
the probabilistic cellular automata models \cite{ghosh2020data},
the models considers stochastic modeling \cite{precharattana2016stochastic},
and the models combine the differential equation and the cellular automata 
\cite{bin2019spread} are proposed.
The models based on the cellular automata considers individual differences 
and can simulate more realistic scenarios. However,
there are limitations in its definition of cell morphology and neighbor rules.

(3) The methods based on the complex network. 
In this approach, the contact between people based on 
the real social networks is considered \cite{draief2006epidemic,pastor2015epidemic}
and thus are able to simulate the epidemic spreading
under scenarios closest to the real-world.
In the complex network, nodes represent individuals and the 
connection between individuals are characterized by links between 
nodes \cite{draief2006epidemic,pastor2015epidemic}.  
To simulate the epidemic spreading, one, for example, has to solve the ordinary differential 
equations on that network \cite{pastor2001epidemic}.
However, to solve the epidemic models on the underlying topology of complex network
is difficult. For example, the Monte Carlo simulations are useless when 
study the network with a large system size due to memory needs \cite{moreno2003epidemic}.
Therefore, unlike the compartment models and the 
cellular automata models, 
the models based on the complex network
focus on the profound impact of the complex properties of real-world networks
on the behavior of
equilibrium and nonequilibrium phenomena \cite{pastor2015epidemic},
or thresholds \cite{castellano2010thresholds,xu2012push} of epidemic spreading.
For example, an epidemic 
threshold that is inversely proportional to the largest
eigenvalue of the connectivity matrix is proposed \cite{boguna2002epidemic} and
Immunization and epidemic threshold of an SIS model in complex
networks is investigated \cite{wu2016immunization,zhang2018epidemic}.
Moreover, various kinds of complex networks are investigated, such as the 
correlated complex networks \cite{moreno2003epidemic} and
the scale-free complex networks \cite{pastor2001epidemic,zhang2018epidemic}.
The models based on the complex network are indeed 
the model closest to the realistic, however, 
it is difficult to obtain a large simulations on those model 
due to the complexity in a given time.

In the realistic social networks, on the one hand,
the random contact, including contacts between people in the public places and
the public transits, becomes the important access for the spreading of infectious 
diseases. On the other hand, the situations are complicated. For example, individuals
vary from one to another: those with good physical condition perhaps become
the asymptomatic carriers and those with strong sense of prevention might become the vaccine recipients
or the mask wearers.
Therefore, in order to obtain a large number of simulations of disease spreading supporting the 
searching of the optimal strategy of disease control, 
a model can simulate the disease spreading under various conditions and quickly return a 
simulation result is needed.

In this paper, a model can efficiently simulate the spreading of infectious 
diseases under random contacts is proposed.
In this approach, the random contact between people is characterized by 
the random matrix with elements randomly generated and the spreading of the diseases 
is simulated by the Markov-like process, a process has slightly 
differences between the Markov process.
We report an interesting property of the proposed model:
the main indicators of the spreading of the diseases such as the death rate are invariant 
of the size of the population. 
Therefore, representative simulations can be conducted on models consist of small number of 
populations.
The main advantage of this model is that it can easily simulate
the spreading of diseases under more realistic scenarios 
and thus is able to give a large number of simulations needed for the searching of the 
optimal control strategy. 
For example, we simulate over $7$ scenarios where the spreading of 
different infectious diseases, 
characterized by their infections and deaths, under different control measures 
are given.
This work is the first work in the series studies aimed at giving the optimal disease control strategy.
Based on this work, the reinforcement learning will be introduced 
to give the optimal control strategy in the following work.

The paper is organized as follows: 
In Sec. 2, we introduce the random matrix and establish the connection
between the random matrix and random contact.
In Sec. 3, the main method is introduced.
In Sec. 4, simulations under different scenarios are conducted. 
Conclusions and discussions are given in Sec. 5.

\section{The main method}
In this section, we introduce the main method, including the approach of simulate the random contact 
by the random matrix and the rule a disease spread on the random matrix.

\subsection{The Random Matrix and the Random Contact}
In this section, we introduce the approach of simulate the random contact 
by the random matrix.

Random matrix is also called probability transition matrix in mathematics, 
which is a basic tool for characterizing Markov processes \cite{2005Random},
It can be used to describe complex systems with random interactions.
For example, physical system in non-equilibrium state \cite{Thomas1998Random}, 
complex communication system \cite{tulino2004random}, and 
complex social network of infectious disease population.
 
\begin{figure}[H]
\centering
\includegraphics[width=0.9\textwidth]{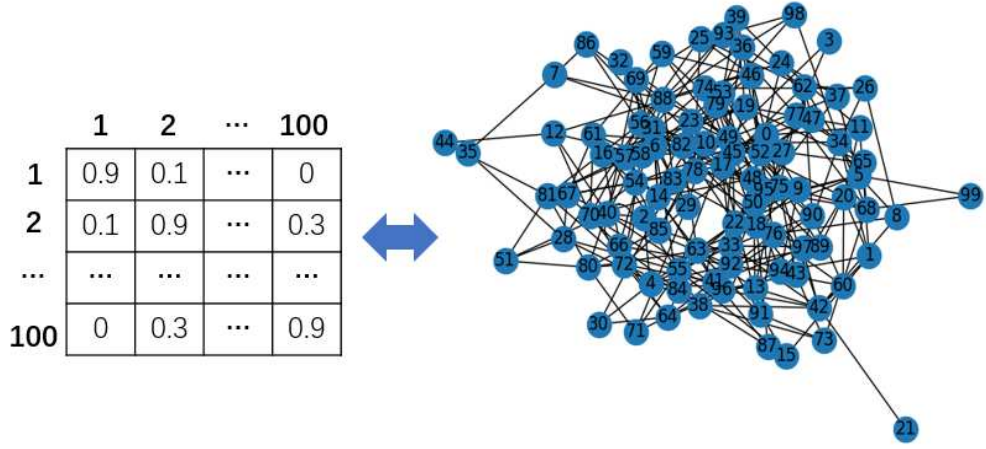}
\caption{An example of simulate the fixed connection and random 
contact between $100$ individuals by the random matrix: the matrix and the corresponding network.
Individual $2$ has closer connection with individual $100$ than $1$. The recover coefficient for each 
individuals is $0.9$. 
}
\label{kk}
\end{figure}

In this approach, we consider a group of individuals indexed from $1$ to $N$.
The social connection of these individuals are characterized by a random matrix $A$ with order $N$.
Here are two basic assumptions of the proposed model.
(1) The element $A_{i,j}$ at $ith$ row and $jth$ column represents the connection between individuals $i$ and $j$.
For example, if there is connections, including the fixed connection and random contact, 
between individuals $i$ and $j$,
the matrix element at position $i$ and $j$ will be assigned with a non-negative value.
Otherwise the element will be $0$. The larger the value, the closer the contact.
For example, a random contact with a stranger at a bus is characterized by $0.01$ and 
a contact with a family member is $0.5$.
In order to simulate both the fixed connection, say the contact between families and colleague,
and random contact, say the contact between strangers at public places, we assign randomly 
generated values, under certain constraint, to the elements of the matrix.
As a result, a symmetric random matrix is generated. 
(2) The element at $ith$ row and $jth$ column also represents the exposure of 
spreading the diseases from the $ith$ individual to the $jth$ individual and vice versa.
Therefore, the diagonal element represent the recover coefficient of the diseases for each individuals. 
For the sake of clarity, we give an example
where $100$ individuals are considered, as shown in Fig. (\ref{kk}). 
In this example, each individual averagely has connection with
$4$ other individuals. The number of contacts for each individual
is rounded from a value sampled from a normal distribution with mean $4$ and variance $4$.
The matrix elements are either $0$ or 
sampled from a normal distribution with mean $0.4$ and variance $0.2$.
For the sake of convenience, the matrix element are named the contact coefficient.
To summarize, the average number of contacts and the average exposure coefficient 
are two constraint applied on the generation of the random matrix revealing the 
social activity.

\subsection{The Rule of Diseases Spreading}
In this section, we introduce the rule where a disease spread on the random matrix.

\begin{figure}[H]
\centering
\includegraphics[width=0.9\textwidth]{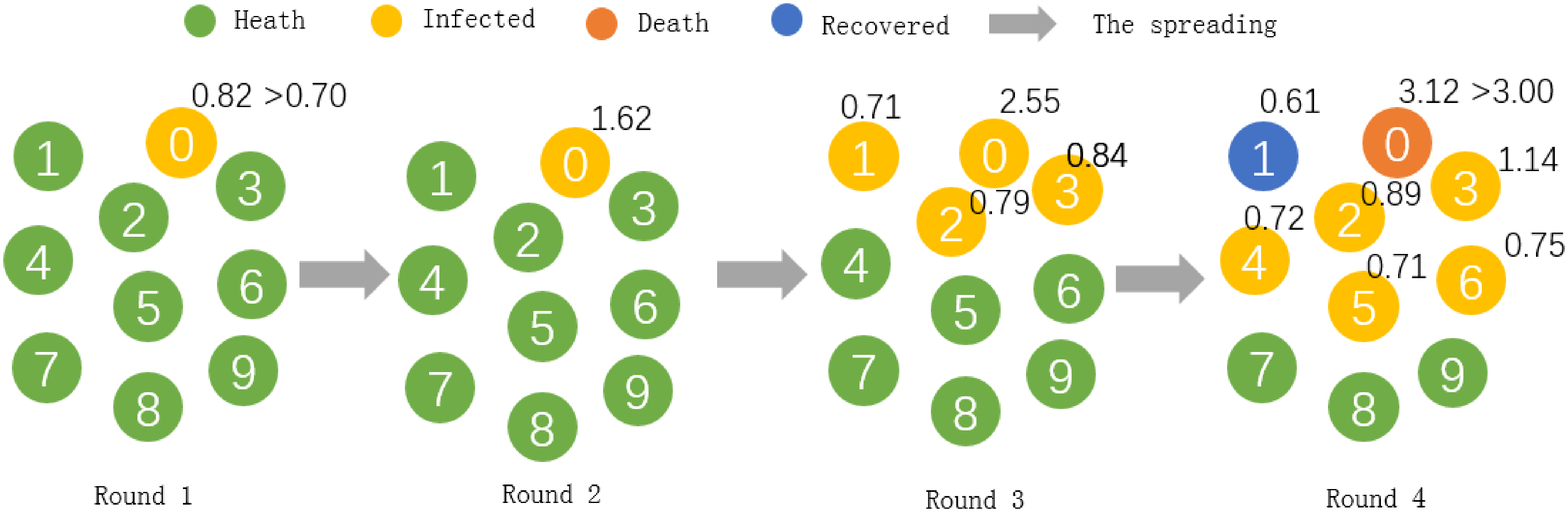}
\caption{An example illustrating the spreading of the disease. 
The pathogenic threshold is $0.70$ and the lethal threshold is $3.00$. }
\label{process_infection}
\end{figure}

After the construction of the random matrix characterizing the social connection between a 
given group of individuals, we set up the following rule to simulate the spreading of the 
diseases.
(1) A vector $x$, named the criterion vector, of size $N\times1$ is given characterizing 
the disease infection status of each individual.
The $ith$ component of the criterion vector evaluates the exposure to the disease.
(2) If the exposure to the disease is below a certain value, the pathogenic threshold, 
the individual is health. Otherwise, he becomes infected. If the exposure
beyond the given value, the lethal threshold, the individual will die. If the exposure
return to a value below the pathogenic threshold, he is recovered.
The diseases are mainly characterized by the recover coefficient and the pathogenic and lethal thresholds.
(3) The spreading of the disease is simulated by considering the random matrix $A$ as the
transition probability matrix and the criterion vector $x$ as the vector of state in a Markov process.
In a transition probability matrix, the summation of a row is $1$ making the total 
probability conserved. However, here, the elements of the matrix is not regarded as a probability
and the summation of the row of the random matrix is not $1$. Therefore, we name it as the 
Markov-like process. The spreading is described by the following equation,
\begin{equation}
x_{n+1}=Ax_{n},%
\end{equation}
where $x_{n+1}$ is the criterion vector of next round.
For the sake of clarity, we give an example where $10$ individuals are considered, 
as shown in Fig. (\ref{process_infection}). 

\begin{figure}[H]
\centering
\includegraphics[width=0.9\textwidth]{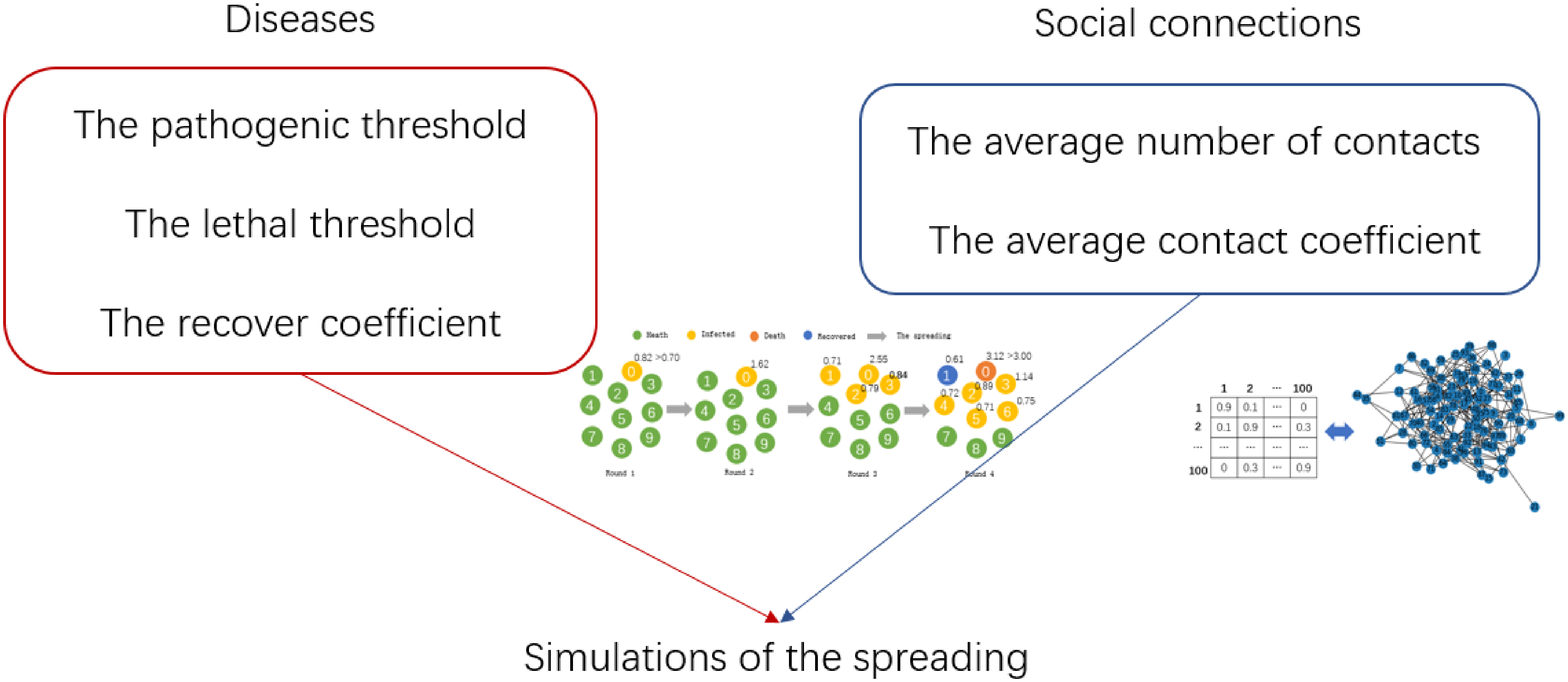}
\caption{An example illustrating the spreading of the disease. 
The pathogenic threshold is $0.70$ and the lethal threshold is $3.00$. }
\label{summary}
\end{figure}

To summarize, in this approach, 
the diseases are characterized by the recover coefficient and the pathogenic and lethal thresholds. And the 
social connections are described by the random matrix generated under the constraint 
characterized by the 
average number of contacts and the average exposure coefficient.
By applying the rules, we can simulate the spreading of different diseases under various realistic situations,
as show in Fig. (\ref{summary}). 

\section{The Model's Parameter and Some Discussions}
In this section, before conducting various simulations 
by the proposed models,
we give explanations to the model's parameters since 
the estimation of model's parameters is always 
important. 
Beside, we report an interesting property of the proposed model and 
give a simple discussion on the outbreak threshold of the epidemic spreading.

\subsection{The Model's Parameter}
Usually, the model's parameters should be estimated and 
verified based on the real data first, and then 
the model could be used to simulate the spreading.
In this work, the main aim is to propose a model that can give a large amount of simulations in a given time 
supporting the searching of the optimal disease control strategy.
Therefore, we focus on testing the flexibility of the proposed model to 
simulate various realistic situations by setting the model's parameters manually.
The estimating of the parameter will be considered in further researches. 

In this section, we consider three types of diseases with low, medium, and high pathogenic and lethal thresholds.
These three types of diseases represent the general and malignant infectious diseases.
Given that the social connections between individuals are described 
by the random matrix generated under the constraint 
characterized by the 
average number of contacts and the average exposure 
coefficient and the matrix will be altered during the simulation
according to different control measures,
we set the 
average number of contacts $4$ and the average exposure coefficient $0.13$.
An overview of the setting of model's parameter is given in Table. (\ref{table1})
 
\begin{table}[H]
\label{table1}
\caption{The setting of model's parameter.
Av-cont-coe, Av-cont-num, Path-thre, Leth-thre, and Rec-coe
stand for the average exposure coefficient, the 
average number of contacts, the pathogenic threshold, the lethal threshold, 
and the recover coefficient respectively.}
\centering
\begin{tabular}{llllllll} 
\hline    
 $\text{Diseases}$ &$\text{Av-exp-coe}$ & $\text{Av-cont-num}$ & $\text{Path-thre}$ & $\text{Leth-thre}$ & $\text{Rec-coe}$ \\
\hline   
$\text{Malignant}$ &$0.13$ & $4.0$ & $0.4$ & $4.0$ & $0.9$ \\   

$\text{Moderate }$ &$0.13$ & $4.0$ & $0.7$ & $6.0$ & $0.7$ \\   

$\text{General}$ &$0.13$ & $4.0$ & $0.9$ & $8.0$ & $0.5$ \\   
\hline
\end{tabular}
\end{table}

\begin{figure}[H]
\centering
\includegraphics[width=0.9\textwidth]{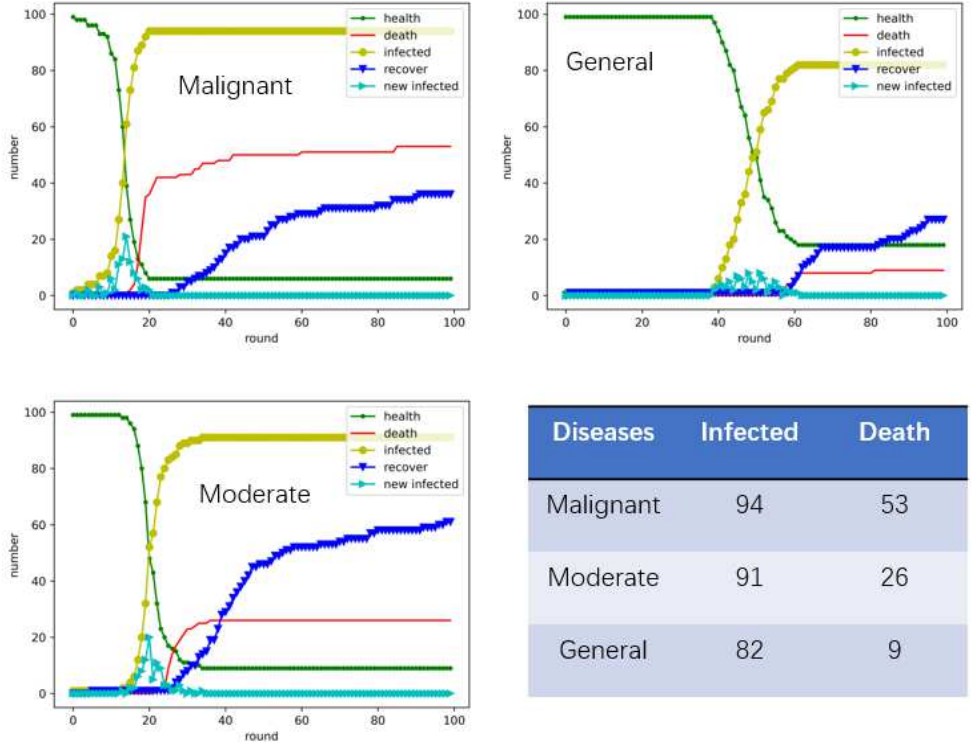}
\caption{A diagram showing the spreading of these 
three types of diseases. We consider $100$ individuals with 
average number of contacts $4$ and the average exposure coefficient $0.13$. }
\label{norm}
\end{figure}

Here is a diagram, Fig. (\ref{norm}), showing the spreading of these 
three types of diseases. The curve trends and 
the infected and death populations show the differences.

\subsection{An Interesting Property: the Model is Invariant of the Size of the Model}
Usually, to simulate the spreading of diseases closer to the realistic situations,
one needs to simulate a system with large number of individuals. 
For example, a small town consists of $100$ thousand populations.
In this section, we report an interesting property of the proposed model: 
the model is invariant of the size.
That is, the main indicators of disease transmission, 
such as the death rate and the infection rate of 
simulations, remain stable and are invariant of the population sizes,
as shown in Table. (\ref{table2}).

To investigate this interesting property, 
we also repeat the experiments on
community networks, a kind of networks describing the social connection of individuals in 
realistic world. The random matrix describing the community networks 
can be easily constructed. For example, for the network proposed in this work,
the elements of the matrix are sampled from a Gauss distribution under given constraint, 
see section 2.1. For the community networks, between individuals in a community, 
the random matrix
is generated in the same way, however, between individuals in different communities,
only selected individuals have connections, say $5$.
Here is the results of another sets of experiments, as shown in Table. (\ref{table3}).

\begin{figure}[H]
\centering
\includegraphics[width=0.9\textwidth]{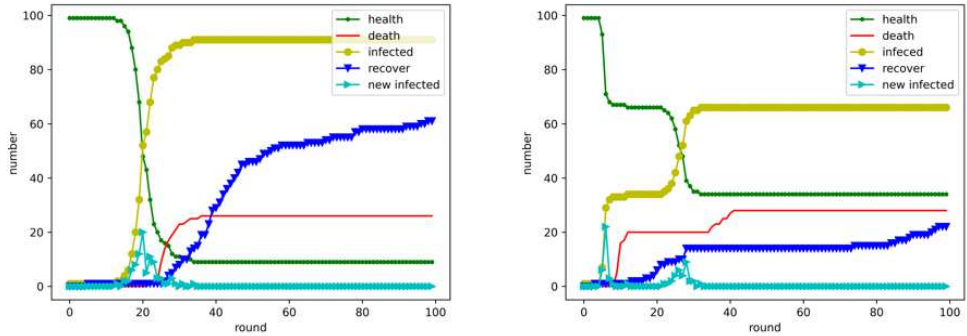}
\caption{A comparision between the simulations of the spreading of diseases on
the network proposed in this work and the community networks.}
\label{compare_norm_comu}
\end{figure}

Tables. (\ref{table2}) and (\ref{table3}) shows that the main 
indicators of disease transmission is invariant of the size of the population.
It means that, without loss of generality,
the following simulations can be conducted in a model of small number 
of population, say $100$,
which saves computer resources to a desirable extent. 
Moreover, Fig. (\ref{compare_norm_comu}) shows the difference between 
simulations of the spreading of diseases on
the network proposed in this work and the community networks.
Although, the construction of those two
random matrices are different, to construct random matrices describing those
networks is easy. It shows in Fig. (\ref{compare_norm_comu}) that the spreading on
the community network can be divided into different stages while that on the 
networks provided in this work has only one stage. It is because the spreading of 
diseases on different communities is not synchronous.

\begin{table}[H]
\label{table2}
\caption{The main indicators of disease transmission in different simulations with different population size.
The average exposure coefficients is $0.13$, the 
average number of contacts is $4$, the pathogenic threshold is $0.7$, the lethal threshold is $6.0$, 
and the recover coefficient is $0.7$.}
\centering
\begin{tabular}{lll} 
\hline    
$\text{N}$ & $\text{Infection rate}$ & $\text{Death rate}$ \\
\hline   
$100$ & $0.910$ & $0.260$  \\   

$500$ & $0.894$ & $0.260$  \\   

$1000$ & $0.895$ & $0.243$  \\   

$2000$ & $0.900$ & $0.249$  \\   
\hline
\end{tabular}
\end{table}

\begin{table}[H]
\label{table3}
\caption{The main indicators of disease transmission in different simulations on community networks.
Con-of-com stands for the number of connections between communities.}
\centering
\begin{tabular}{llll} 
\hline    
$\text{N}$ & $\text{Con-of-com}$ & $\text{Infection rate}$ & $\text{Death rate}$ \\
\hline   
$100$& $1$ & $0.660$ & $0.280$  \\   

$500$& $1$ & $0.664$ & $0.286$  \\   

$1000$& $1$ & $0.672$ & $0.314$  \\   

$2000$& $1$ & $0.665$ & $0.297$  \\   

$2000$& $2$ & $0.699$ & $0.305$  \\   

$2000$& $5$ & $0.686$ & $0.301$  \\   

$2000$& $10$ & $0.747$ & $0.309$  \\   
\hline
\end{tabular}
\end{table}

In the following simulations, 
we consider $100$ individuals.
Here we only report this interesting property and further research on this phenomenon 
will be carried out in later works.

\subsection{A Simple Discussion on Whether the Infectious Disease Will Break Our or Not}
In this section, we investigate the outbreak threshold of the epidemic spreading.
Instead of a rigorous discussion, we give a simple discussion on whether the infectious 
disease will outbreaks or not. Further investigation on the 
equilibrium and nonequilibrium phenomena and thresholds 
of epidemic spreading
will be given in later work.

\begin{figure}[H]
\centering
\includegraphics[width=0.9\textwidth]{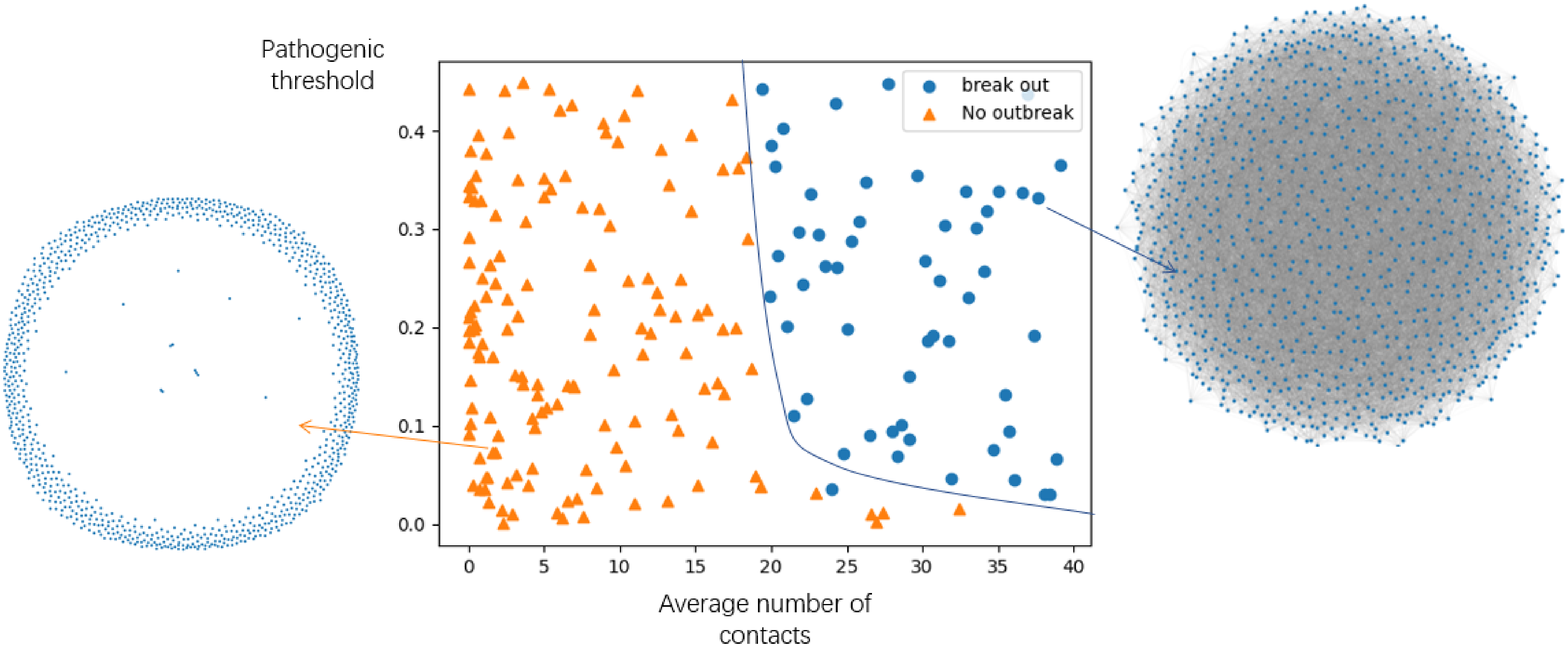}
\caption{A diagram showing the result of simulations where 
the spreading of different diseases
on various situations are considered. The orange triangle represents that 
the epidemic dose not break out and the blue dot represents breakout. 
The recover coefficient is $0.7$ and the average exposure rate is $0.13$.}
\label{breakout}
\end{figure}

To explore the outbreak threshold of the epidemic spreading, 
in this approach, we take the advantages of 
the proposed method and obtain 
a large number of simulations where the spreading of different diseases
on various situations are considered. Here, the outbreak of the epidemic spreading
is defined as over $70\%$ populations are infected. 
It shows in Fig. (\ref{breakout})
that there is a clear boundary between the outbreak and no outbreak cases.
For example, the diseases with low pathogenic threshold won't break out regardless of the 
average number of contacts of the social network.

\section{The Simulation of the Spreading of Different Diseases under Different Realistic Scenarios}
In this section, we simulation of the spreading of different diseases under different realistic scenarios.
The diseases are mainly characterized by the pathogenic and lethal thresholds and the realistic scenarios
are simulated by the random matrix.
For instance, to simulate the spreading of the diseases under the scenario where
the control measures of isolating the infected is adopted, the random matrix is altered after each round.
The elements corresponding to the infected individuals are all set to $0$.

\subsection{Scenarios 1: the Effectiveness of Passive Quarantine}
\begin{figure}[H]
\centering
\includegraphics[width=1.0\textwidth]{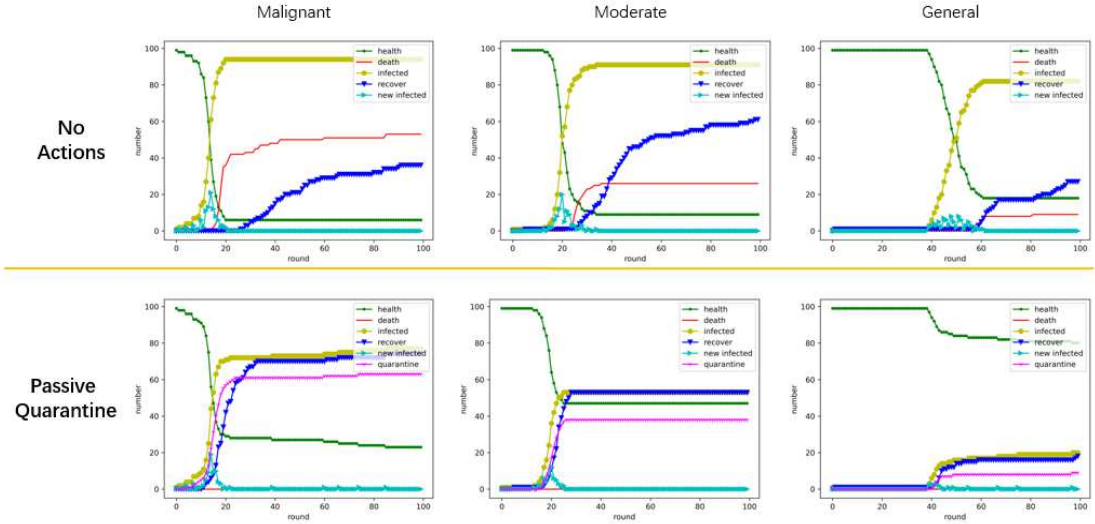}
\caption{The simulations of spreading under passive quarantine and no actions.
The curve trends show the effectiveness of passive quarantine.
For example, the infected and death are obviously reduced.}
\label{passive_quarantine}
\end{figure}

In this section, we consider the effectiveness of passive quarantine.
The quarantine can be simply simulated by setting the 
matrix elements corresponding to the individual $0$ if we 
find this individual need to be in quarantine. 
In this simulation 
only those who go to hospital seeking 
for a treatment themselves
are considered to be in quarantine. 
Usually, not all infected individuals go to hospital seeking for medical treatment
and not all be diagnosed due to wrong diagnosis. Therefore, in this simulation,
we set the probability of being diagnosed $0.3$. That is for those 
who are infected, only $30\%$ will be diagnosed and quarantined.
It shows in Fig. (\ref{passive_quarantine}) and Table. (\ref{table4}) that the 
quarantine contributes greatly to reduce the infected and death cases 
in the spreading of the diseases.

\begin{table}[H]
\label{table4}
\caption{The main indicators of spreading under passive quarantine and no actions.
The infection rate and the death rate (the ratio between death and all population) 
show the effectiveness of passive quarantine.}
\centering
\begin{tabular}{llll} 
\hline    
$\text{Diseases}$ & $\text{Scenarios}$ & $\text{Infected}$ & $\text{Death}$ \\
\hline   
$Malignant$& $No quarantine$ & $94$ & $53$  \\   
$ $& $Quarantine$ & $77$ & $0$  \\ 
$Moderate$& $No quarantine$ & $91$ & $26$  \\ 
$ $& $Quarantine$ & $53$ & $0$  \\ 
$General$& $No quarantine$ & $82$ & $9$  \\ 
$ $& $Quarantine$ & $20$ & $0$  \\ 
\hline
\end{tabular}
\end{table}

\begin{figure}[H]
\centering
\includegraphics[width=0.9\textwidth]{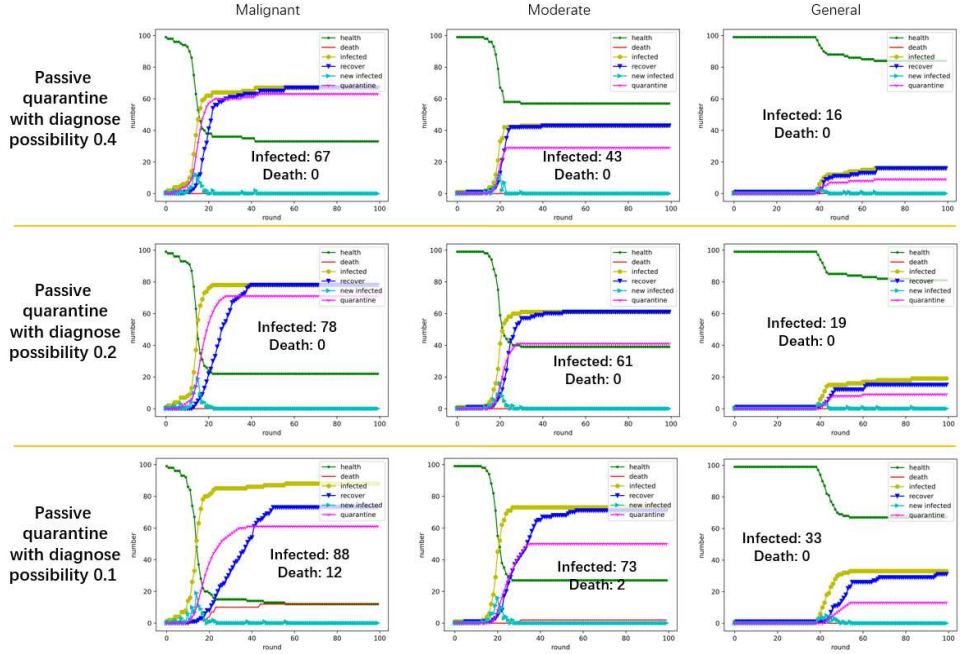}
\caption{The simulations of spreading under passive quarantine with different 
diagnose probabilities.}
\label{passive_quarantine2}
\end{figure}

Here are another simulations where the probability of being diagnosed $0.4$, $0.2$, 
and $0.1$ respectively,
as shown in Fig. (\ref{passive_quarantine2}).
The simulations suggest that the probability of being diagnosed
to be an infected is crucial for the deaths. For example, when the probability
is below $0.1$, there are deaths.

\subsection{Scenarios 2: the Effectiveness of Active Quarantine }
\begin{figure}[H]
\centering
\includegraphics[width=1.0\textwidth]{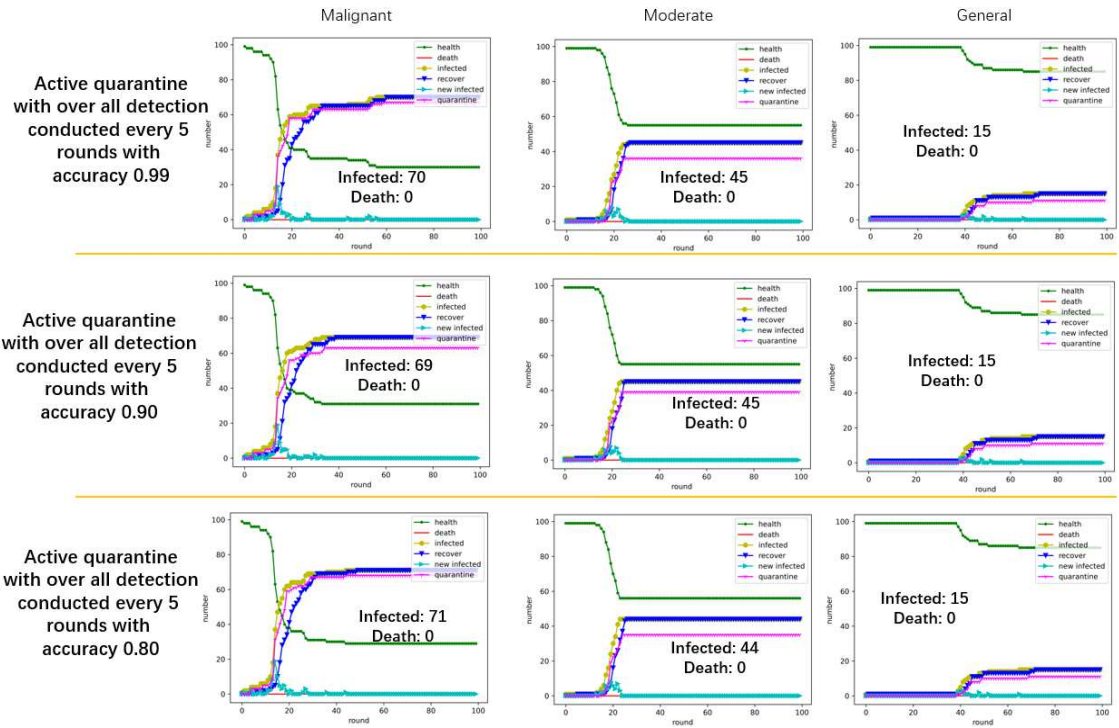}
\caption{The simulations of spreading under active quarantine with different 
overall disease detection accuracies. The overall disease detection 
frequency is every $5$ round.}
\label{active_quarantine}
\end{figure}

\begin{figure}[H]
\centering
\includegraphics[width=1.0\textwidth]{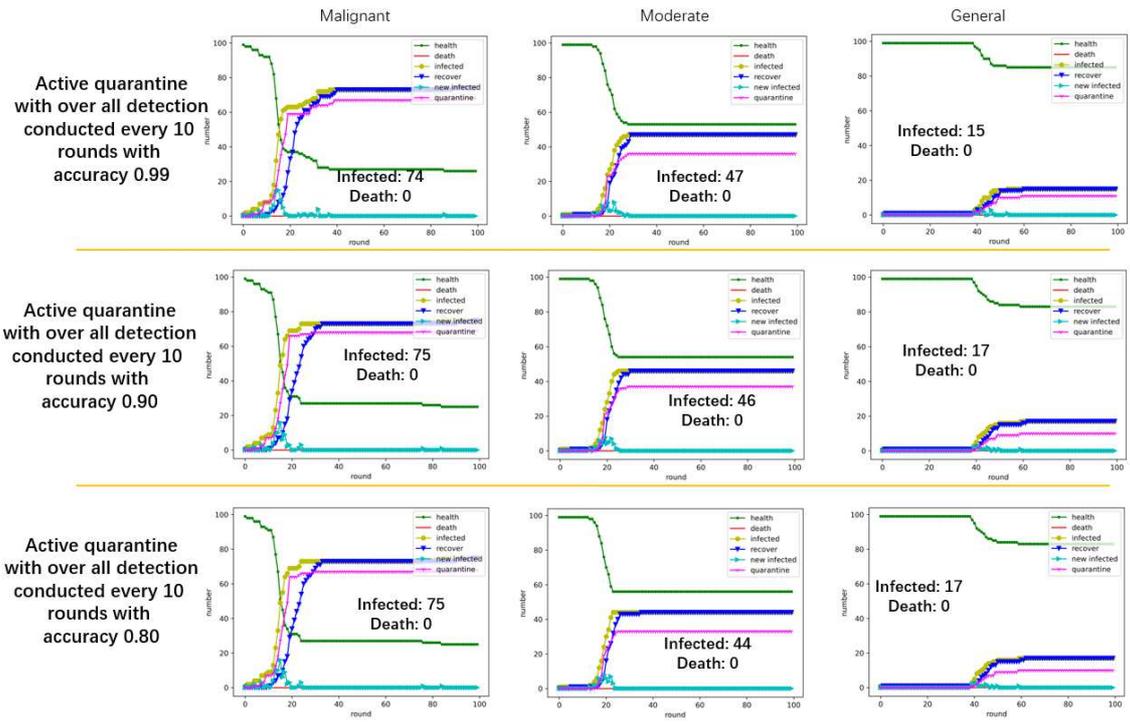}
\caption{The simulations of spreading under active quarantine with different 
overall disease detection accuracies. The overall disease detection 
frequency is every $10$ round.}
\label{active_quarantine2}
\end{figure}

In this section, we consider the effectiveness of active quarantine.
Unlike the passive quarantine, in the active quarantine, we 
conduct a disease detection for all individuals and those found to be 
infected are in quarantine. The active quarantine considers those 
who don't go to hospital for treatment themselves.
Given that the method we use for disease detection is imperfect,
that is, the method can't find all infected individuals. We simulate 
the cases where the detection recall is $99\%$, $90\%$, and $80\%$
respectively. In realistic life, conducting an overall disease detection
costs considerable human and financial costs, therefore, in the simulation,
the overall disease detection is conducted every $5$ and $10$ rounds.
Figs. (\ref{active_quarantine}) and (\ref{active_quarantine2}) show 
the effectiveness of active quarantine.
The simulations of proposed model show that, 
the overall disease detection helps little 
with the control of spreading of the diseases.
Increasing the frequency and accuracy of the overall disease detection
contributs little to the control of spreading of the diseases too.

\subsection{Scenarios 3: the Effectiveness of the Green Code}
\begin{figure}[H]
\centering
\includegraphics[width=1.0\textwidth]{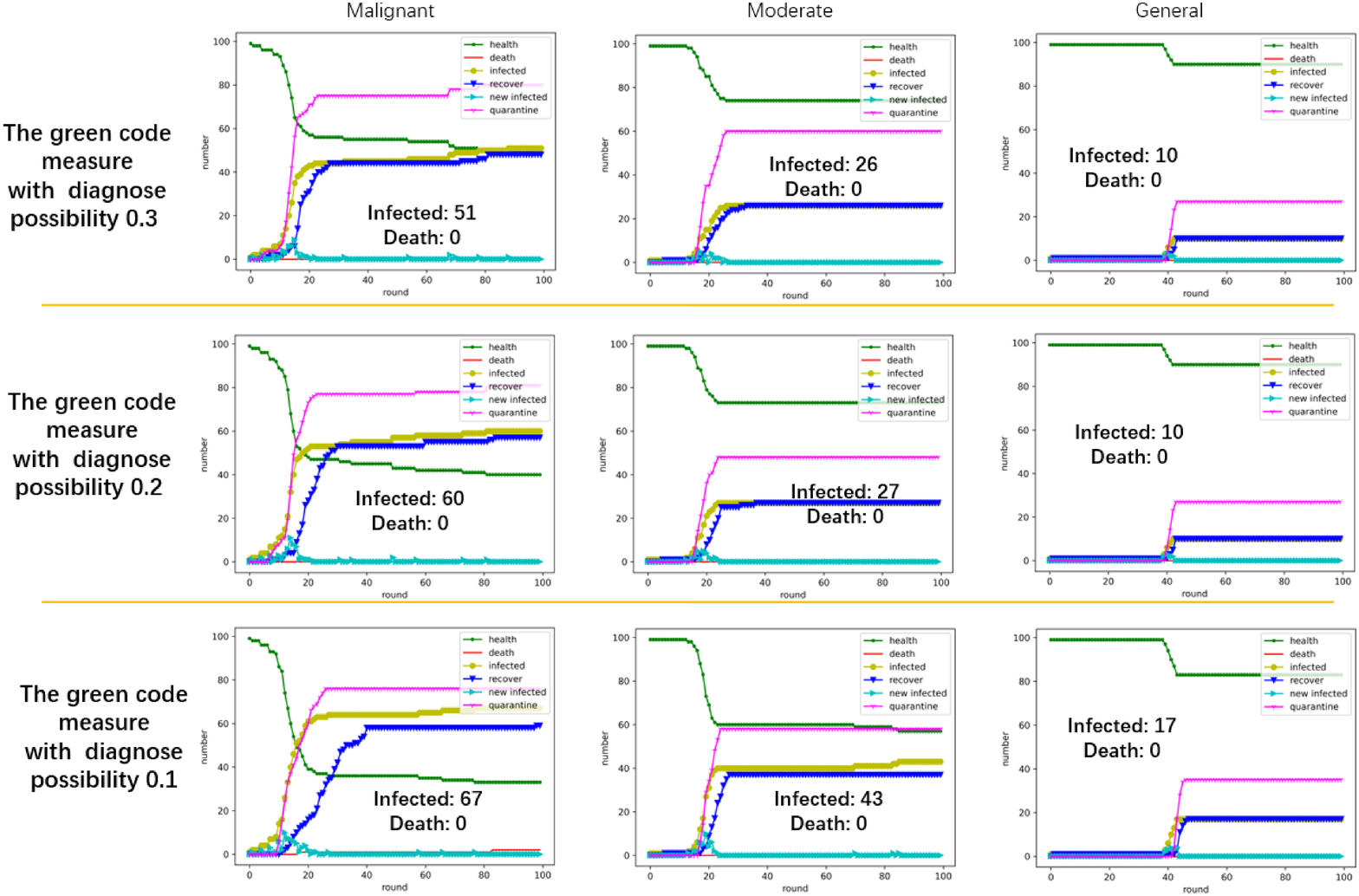}
\caption{The simulations of spreading under the green code measure with different 
diagnose probabilities.}
\label{luma}
\end{figure}

\begin{figure}[H]
\centering
\includegraphics[width=1.0\textwidth]{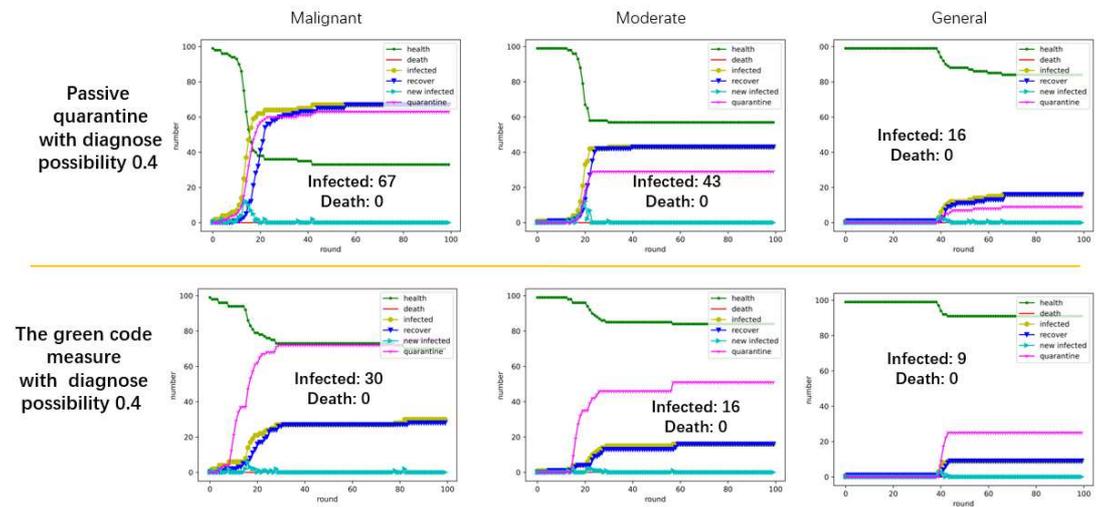}
\caption{The simulations of spreading under the passive quarantine with and without
green code measure.}
\label{luma2}
\end{figure}

In this section, we consider the effectiveness of the green code,
a measure used in China to control the COVID-19.
In this control measure, not only those who are diagnosed are quarantined but also 
those who have direct contact with the infected, namely close contacts, are quarantined 
too.
In this simulations, those who go to hospital seeking for a medical treatment
themselves are possibly be diagnosed as infected.
The result of simulations are shown in Figs. (\ref{luma}) and (\ref{luma2}).
It shows that the green code measure contributes to the control of the spreading of 
the diseases obviously. Especially from the curve trends in  Fig. (\ref{luma2}),
the infected population is reduced obviously. 

\subsection{Scenarios 4: the Effectiveness of the Green Code and Overall Detection}
In this section, we consider the effectiveness of the green code and overall detection.
That is, the overall detection is conducted to find the infected population such as those 
who never go to hospital. Moreover, the infected and the close contacts are quarantined.

\begin{figure}[H]
\centering
\includegraphics[width=1.0\textwidth]{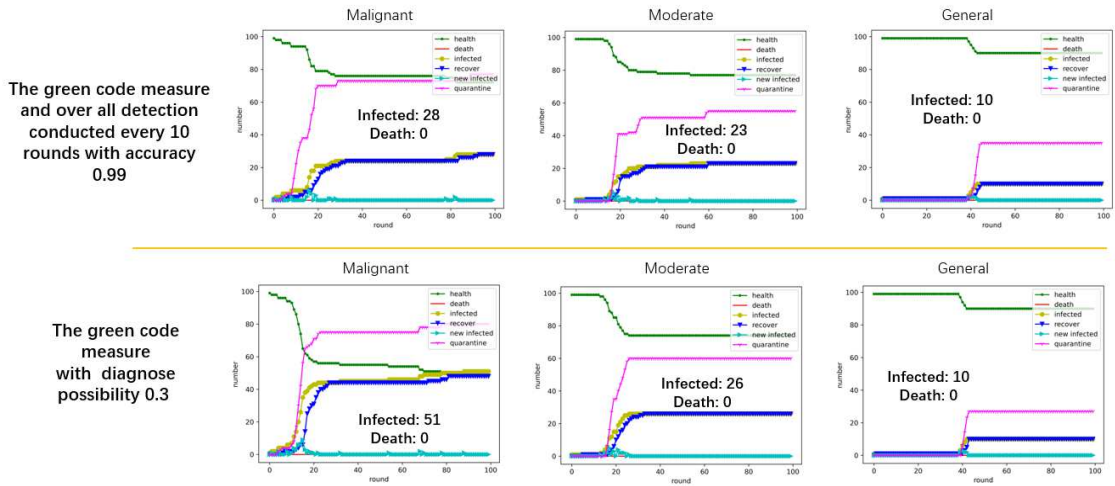}
\caption{The simulations of spreading under the green code measure with and without overall 
detection. }
\label{luma_overall}
\end{figure}

The result of simulations are shown in Fig. (\ref{luma_overall}).
It shows that the green code measure and overall 
detection together contribute to the control of the spreading of 
the diseases to some degree. For the malignant infectious disease
the control measure is with higher effectiveness.
The simulation suggests that to effectively control the spreading of the disease,
to find the infected and their close contacts and make them quarantined are the 
keys.

\subsection{Scenarios 5: Asymptomatic Carriers}
In this section, we consider the asymptomatic carriers.
The asymptomatic carriers are individuals who are infected and are contagious
but have no uncomfortable signs and symptoms.
Usually, the asymptomatic carriers are hard to be found.
In the following simulations, we consider 
the spreading of diseases with asymptomatic carriers 
under the most strict control measures, the green code and overall detection.
Here, we make the assumption that the asymptomatic carriers 
can not be found even in the overall detection.

\begin{figure}[H]
\centering
\includegraphics[width=1.0\textwidth]{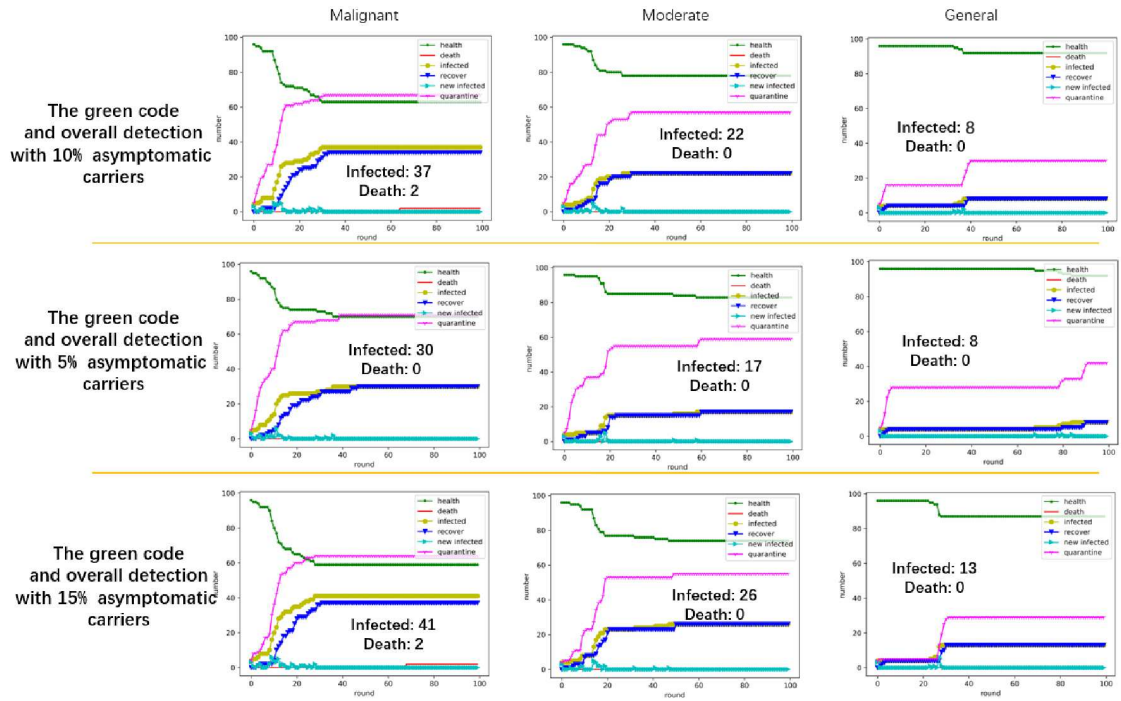}
\caption{The simulations of spreading with asymptomatic carriers
 under the green code and overall 
detection. }
\label{AC_v2}
\end{figure}

The result of simulations are shown in Fig. (\ref{AC_v2}).
It shows that the asymptomatic carriers are the main group 
to break down the strict control measure, the green code and overall detection.
Especially when there are over $10\%$ of the populations are 
asymptomatic carriers, there are deaths. 

\subsection{Scenarios 6: Vaccination}
In this section, we consider the spreading of diseases with vaccination.
The vaccination is a useful method to control the spreading of diseases.
In this simulation, we make the assumption that the vaccination
will increase the resistance of the pathogens. That is the 
pathogenic and lethal thresholds stay the same, but the recover coefficient 
will the lowered.
The lower the recover coefficient the quicker the infected will recover.
Here, after vaccination, the recover coefficient multiplies $0.7$.

\begin{figure}[H]
\centering
\includegraphics[width=1.0\textwidth]{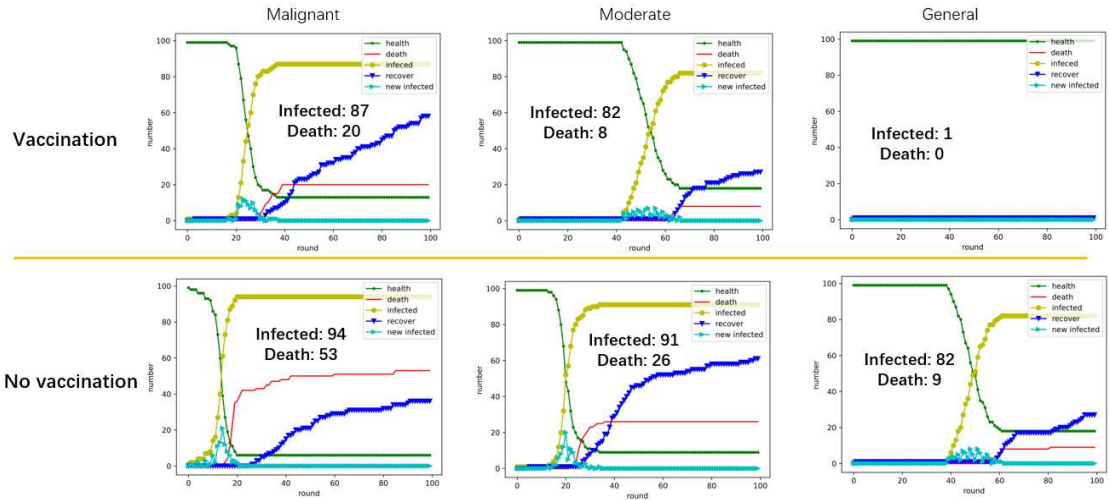}
\caption{The simulations of spreading with and without vaccination. }
\label{vaccination}
\end{figure}

The result of simulations are shown in Fig. (\ref{vaccination}).
It shows that the vaccination can effectively control the spreading 
of diseases, especially for the general infectious diseases.
However, to avoid deaths, only vaccination is not enough.
That is, we need control measures such as the quarantine and green code.

\subsection{Scenarios 7: Mask Wears}
In this section, we consider diseases spread by air,
and to wear a mask can efficiently prevent the spreading.
Here, we consider two kinds of mask, one is medical mask and the other is
the general mask.
In this simulation, the medical mask can reduce the exposure coefficient by multiplying 
$0.9$ and the general mask $0.95$.
Without loss of generality, we make assumption that all individuals 
wear masks.

\begin{figure}[H]
\centering
\includegraphics[width=1.0\textwidth]{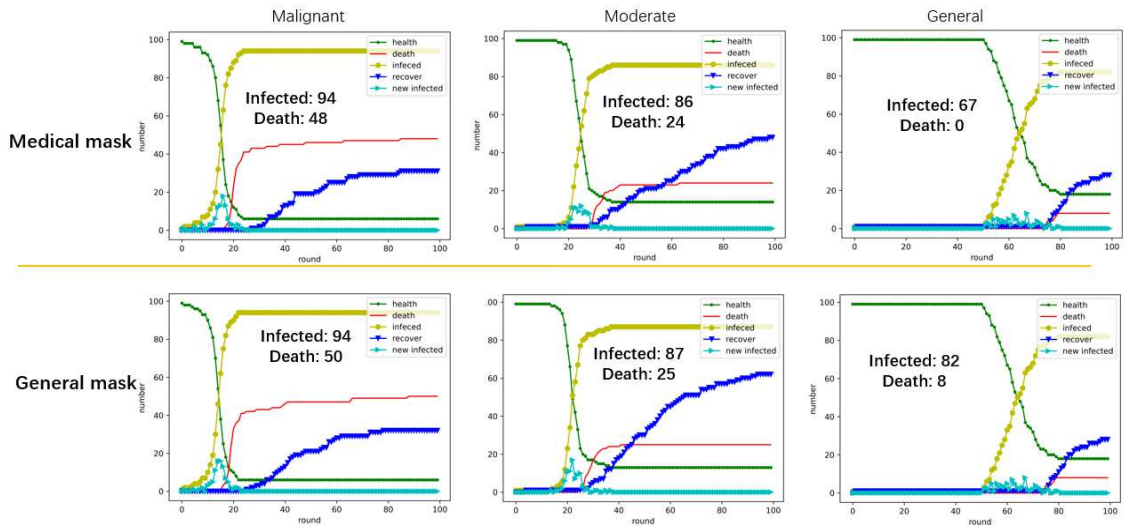}
\caption{The simulations of spreading with medical and general masks. }
\label{mask}
\end{figure}

The result of simulations are shown in Fig. (\ref{mask}).
It shows that wearing a mask can control the spreading 
of diseases, especially for the general infectious diseases
the medical mask reduces the deaths to $0$.

\section{Conclusions and Discussions}

Establishing a model that can simulate the spread of infectious diseases is the key to studying 
the law of infectious disease transmission and predicting the trend of transmission, 
and it is of great significance to control the spread of infectious diseases. 
To find the optimal control strategy, we need a large number of simulations to support the 
searching process. Therefore, one needs the model that can at one hand simulate the spreading 
of diseases under realistic conditions and at the other hand quickly return the result 
of the simulations.
In the realistic social network, on the one hand, the random contact, including the contact with 
strangers at public places, becomes more and more important access for the spreading of 
diseases. One the other hand, the individuals are different form one to the other, 
and the individual differences
will effect the spreading of diseases. For example, the asymptomatic carriers will cause a wider spread of the disease.

In this work, a model that can simulate the spreading of diseases under realistic situations with  
random contact is proposed. Unlike the conventional approaches, such as the differential equation based compartment models,
the cellular automata and the complex network based models,
the social connections, in this approach, are described 
by using the random matrix
which is generated under given constraint, the average contact population and the average exposure coefficient.
The spreading of the disease is simulated by using the Markov-like process, where
the random matrix is considered as a probability transition matrix with the exposure coefficient (the elements
of the random matrix) the transition probability.
The diseases are characterized by the recover coefficient
and the pathogenic and lethal thresholds.
Instead of estimating the model's parameters according to the real data, we 
firstly manually choose the parameters, in order to simulate the spreading of diseases
under various realistic scenarios.
We report an interesting property of the proposed method: the major indicators such as the infection and death rates
are almost invariant of the size of the model.

A set of experiments show that the proposed model can 
efficiently simulate the spreading of various diseases under various realistic situations.
Therefore, the proposed model can give a large number of simulations supporting the searching of 
optimal diseases control strategy in the following researches.
In the next research, the reinforcement learning method, a method that is good at finding the optimal strategy,
will be introduced. Together with the proposed model, we give a further investigation to the optimal 
control strategy of the infectious diseases.

\section{Acknowledgments} 
We are very indebted to Prof. Wu-Sheng Dai for his enlightenment and encouragement.
We are very indebted to Prof. Guan-Wen Fang for his encouragement. 
This work is supported by National Natural Science Funds of China (Grant No. 62106033), Yunnan Youth
Basic Research Projects (202001AU070020), and Doctoral Programs of Dali University
(KYBS201910).
YuanYuan Wang acknowledges the support from
Yunnan Fundamental Research Projects (grant NO. 2019FD099)

\section{Conflict of interest statement} 
We declare that we have no financial and personal relationships with other people or organizations 
that can inappropriately influence our work, there is no professional or other
personal interest of any nature or kind in any product, service and/or company that could
be construed as influencing the position presented in, or the review of, the present manuscript.











\end{document}